\documentclass[12pt,a4paper]{article}
\usepackage{epsfig}
\begin{document}
\title{Production of the top-pions at the THERA collider\\ based $\gamma p$ collisions}

\author{Chongxing Yue $^{a}$, Hongjie Zong $^{b}$, Shunzhi Wang$^{b}$\\
{\small $^{a}$ Department of Physics , Liaoning Normal University,
Dalian 116029, China
\thanks{E-mail:cxyue@lnnu.edu.cn}}\\
{\small $^{b}$ College of Physics and Information Engineering,}\\
\small{ Henan Normal University, Henan 453002, China } \\ }

\date{\today}

\maketitle

\begin{abstract}
 In the framework of the topcolor-assisted technicolor (TC2)
models, we study the production of the top-pions $\pi^{0}_{t}$,
$\pi_{t}^{\pm}$ via the processes $ep\rightarrow\gamma
c\rightarrow\pi^{0}_{t}c$ and $ep\rightarrow\gamma
c\rightarrow\pi^{\pm}_{t}b$ mediated by the anomalous top coupling
$tc\gamma$. We find that the production cross section of the
process $ep\rightarrow\gamma c\rightarrow\pi^{0}_{t}c$ is very
small.  With reasonable values of the parameters in TC2 models,
the production cross section of the process $ep\rightarrow\gamma
c\rightarrow\pi^{\pm}_{t}b$ can reach $ 1.2pb$. The charged
top-pions $\pi^{\pm}_{t}$ might be directly observed via this
process at the THERA collider based $\gamma p$ collisions.

\end{abstract}

\vspace{2.0cm} \noindent

{\bf PACS number(s)}: 12.60Nz,14.80.Mz,12.15.Lk,14.65.Ha

\newpage

\vspace{.5cm}

To completely avoid the problems arising from the elementary Higgs
field in the standard model (SM), various kinds of dynamical
electroweak symmetry breaking (EWSB) models have been proposed,
and among which the topcolor scenario is attractive because it
provides an additional source of EWSB and solves heavy top quark
problem. Topcolor-assisted technicolor (TC2) models $^{[1]}$,
flavor-universal TC2 models $^{[2]}$, top see-saw models $^{[3]}$,
and top flavor see-saw models $^{[4]}$ are four of such examples.
The common feature of such type of models is that the topcolor
interactions are assumed to be chiral critically strong at the
scale $1 TeV$, and it is coupled preferentially to the third
generation. EWSB is mainly generated by TC interactions or other
strong interactions. The topcolor interactions also make small
contributions to EWSB and give rise to the main part of the top
quark mass $(1-\epsilon) m_{t}$ with $0.03\leq \epsilon \leq 0.1$.
Then, the presence of the physical top-pions in the low-energy
spectrum is an inevitable feature of these models. Thus, studying
the production of the top-pions at present and future high energy
colliders can help the high-energy experiments to search for
top-pions, test topcolor scenario and further to probe EWSB
mechanism.

The production and decay of the technipions predicted by the
technicolor sector have been extensively studied in the literature
$^{[5,6]}$. Combing resonant and non-resonant contributions, the
signals of the technipions are recently studied at the lepton
colliders and the hadron colliders $^{[7]}$. The production and
decays of the top-pions at the lepton colliders and the hadron
colliders are studied in several instances $^{[8,9,10]}$.

For TC2 models, the underlying interactions, topcolor
interactions, are non-universal, and therefore do not posses a GIM
mechanism. This is another feature of this kind of models due to
the need to single out the top quark for condensation. The
non-universal gauge interactions result in the flavor changing
neutral current (FCNC) vertices when one writes the interactions
in the quark mass eigenbasis. The top-pions have large Yukawa
coupling to the third family fermions and can induce the new FC
couplings, which generate the large anomalous top couplings $tcv
(v=\gamma, Z, or g)^{[11]}$. Thus, the top-pions $\pi^{0}_{t}$,
$\pi^{\pm}_{t}$ can be produced via the processes $\gamma
c\rightarrow t\rightarrow \pi^{0}_{t}c$ and $\gamma c\rightarrow
t\rightarrow \pi^{\pm}_{t}b$. Our results show that the production
rate of the neutral top-pion $\pi^{0}_{t}$ is very small and
$\pi^{0}_{t}$ can not be detected at the THERA collider based
$\gamma p$ collisions via the process $ep\rightarrow\gamma
c\rightarrow \pi^{0}_{t}c$. For the process $ep\rightarrow\gamma
c\rightarrow \pi^{\pm}_{t}b$, we find that several tens and up to
thousand events of the charged top-pions $\pi^{\pm}_{t}$ can be
produced per year by assuming the integrated luminosity $L=750
pb^{-1}$ and the center-of-mass energy $\sqrt{s}=1000GeV$ for the
THERA collider based $\gamma p$ collisions $^{[12]}$. The charged
top-pions $\pi^{\pm}_{t}$ may be observed at the THERA collider.

As it is well known, the couplings of the top-pions to the three
family fermions are non-universal. The top-pions have large Yukawa
couplings to the third generation and can induce large flavor
changing couplings. The couplings of the top-pions $\pi^{0}_{t}$,
$\pi^{\pm}_{t}$ to quarks can be written as $^{[1,8]}$:
\begin{eqnarray}
&&\frac{m_{t}}{\sqrt{2}F_{t}}\frac{\sqrt{\nu^{2}_{w}-F^{2}_{t}}}{\nu_{w}}[iK^{tt}_{UR}K^{tt*}_{UL}
\overline{t_{L}}t_{R}\pi^{0}_{t}
+\sqrt{2}K^{tt*}_{UR}K^{bb}_{DL}\overline{t_{R}}b_{L}\pi^{+}_{t}+iK^{tc}_{UR}K^{tt*}_{UL}
\overline{t_{L}}c_{R}\pi^{0}_{t} \hspace{1cm}\nonumber\\
&&\hspace{0.5cm}+\sqrt{2}K^{tc*}_{UR}K^{bb}_{DL}\overline{c_{R}}b_{L}\pi^{+}_{t}+h.c.],
\end{eqnarray}
 where $F_{t}=50 GeV$ is the top-pion decay constant and $\nu_{w}=\nu/\sqrt{2}=174 GeV$. It has been shown
 that the values of the coupling parameters can be taken as:

$$K^{tt}_{UL}=K^{bb}_{DL}=1, \hspace{1cm} K^{tt}_{UR}=1-\epsilon,\hspace{1cm}
K^{tc}_{UR}\leq\sqrt{2\epsilon-\epsilon^{2}},$$ with a
model-dependent parameter $\epsilon$. In the following
calculation, we will take
$K^{tc}_{UR}=\sqrt{2\epsilon-\epsilon^{2}}$ and take $\epsilon$ as
a free parameter.

The neutral top-pion $\pi^{0}_{t}$ and the charged top-pions
$\pi^{\pm}_{t}$ can generate the anomalous top quark couplings
$tcv(v=\gamma, Z, or g)$ via the tree-level FC couplings
$\pi^{0}_{t}tc$ and $\pi^{\pm}_{t}bc$, respectively. However,
compared the contributions of $\pi^{0}_{t}$ to the couplings
$tcv$, the contributions of $\pi^{\pm}_{t}$ to the couplings $tcv$
are very small and can be safely ignored. The effective form of
the anomalous top quark coupling vertex $t-c-\gamma$, which arises
from the tree-level FC coupling $\pi_{t}^{0}\bar{t}c$, can be
written as$^{[11]}$:
\begin{eqnarray}
&&\Lambda^{\mu}_{tc\gamma}=ie[\gamma^{\mu}F_{1\gamma}+p_{t}^{\mu}F_{2\gamma}+p_{c}^{\mu}F_{3\gamma}]
\end{eqnarray}
where
\begin{eqnarray*}
&&F_{1\gamma}=\frac{2A}{3}[B_{0}+m^{2}_{\pi_{t}}C_{0}-2C_{24}+m^{2}_{t}(C_{11}-C_{12})
-B^{*}_{0}-B^{\prime}_{1}],\nonumber\\
&&F_{2\gamma}=\frac{4m_{t}A}{3}[C_{21}+C_{22}-2C_{23}],\nonumber\\
&&F_{3\gamma}=\frac{4m_{t}A}{3}[C_{22}-C_{23}+C_{12}],\nonumber
\end{eqnarray*}
with
$$A=\frac{1}{16\pi^{2}}[\frac{m_{t}}{\sqrt{2}F_{t}}\frac{\sqrt{\nu^{2}_{w}-F^{2}_{t}}}{\nu_{w}}]^{2}
K^{tc}_{UR}K^{tt*}_{UL}.$$
The expressions of two- and three-point scalar integrals $B_{n}$
and $C_{ij}$ are $^{[13]}$:
\begin{eqnarray*}
&&B_{n}=B_{n}(-\sqrt{\hat{s}},m_{t},m_{t}),\hspace{9.5cm}\\
&&B^{*}_{n}=B_{n}(-p_{c},m_{\pi_{t}},m_{t}),\\
&&B^{\prime}_{n}=B_{n}(-p_{t},m_{\pi_{t}},m_{t}),\\
&&C_{ij}=C_{ij}(p_{t},-\sqrt{\hat{s}},m_{\pi_{t}},m_{t},m_{t}),\\
&&C_{0}=C_{0}(p_{t},-\sqrt{\hat{s}},m_{\pi_{t}},m_{t},m_{t}).
\end{eqnarray*}

 Ref.[11] has shown that the anomalous top quark coupling $tc\gamma$
 can give significant contributions to the rare top
 decay $ t \rightarrow c \gamma $ and single top production via
 the process $ e^{+} e^{-} \rightarrow \overline{t} c $. For
 instance, the value of the branching ratio $ Br(t \rightarrow c
 \gamma )$ varies between $7.9 \times 10^{-7} $ and $ 4.6 \times
 10^{-6}$ for $ m_{\pi_{t}} = 300GeV$ and the parameter $\epsilon
 $ in the range of $ 0.01--0.08 $, which can approach the
 corresponding experimental threshold. In this letter, we study
 the contributions of this anomalous top quark coupling $tc\gamma$ to
 the production of the top-pions in the THERA collider based $
 \gamma p $ collisions.

Ref.[1] has estimated the mass of the top-pions in the fermion
loop approximation and given $180GeV\leq m_{\pi_{t}}\leq 240GeV$
for $m_{t}=175GeV$ and $0.03\leq \epsilon\leq 0.1$. The limits on
the mass of the top-pion may be obtained via studying its effects
on various experimental observables. For example, Ref.[14] has
shown that the process $b\rightarrow s \gamma$, $B- \bar{B}$
mixing and $D- \bar{D}$ mixing demand that the top-pion is likely
to be light, with mass of the order of a few hundred GeV. Since
the negative top-pion corrections to the $Z\rightarrow b\bar{b}$
branching ratio $R_{b}$ become smaller when the top-pion is
heavier, the precise measurement value of $R_{b}$ gives rise to a
certain lower bound on the top-pion mass $^{[15]}$. It was shown
that the top-pion mass should not be lighter than the order of $1
TeV$ to make TC2 models consist with the LEP/SLD data $^{[16]}$.
We restudied the problem in Ref.[17] and find that the top-pion
mass $m_{\pi_{t}}$ is allowed to be in the range of a few hundred
GeV depending on the models. Thus, the value of the top-pion mass
$m_{\pi_{t}}$ remains subject to large uncertainty $^{[18]}$.
Furthermore, Ref.[8] has shown that the top-pion mass can be
explored  up to $300-350 GeV$ via the process $p\bar{p}\rightarrow
\pi_{t}^{0}\rightarrow \bar{t}c$ and $p\bar{p}\rightarrow
\pi_{t}^{\pm}x$ at the Tevatron and LHC. Thus, we will take
$m_{\pi_{t}}$ as a free parameter and assume it to vary in the
range of $200GeV-450GeV$ in this letter. In this case, the
dominant decay modes of the charged top-pions $\pi_{t}^{\pm}$ are
$\bar{t}b$ or $t\bar{b}$.

\begin{figure}[htb]
\vspace*{-8cm}
\begin{center}
\epsfig{file=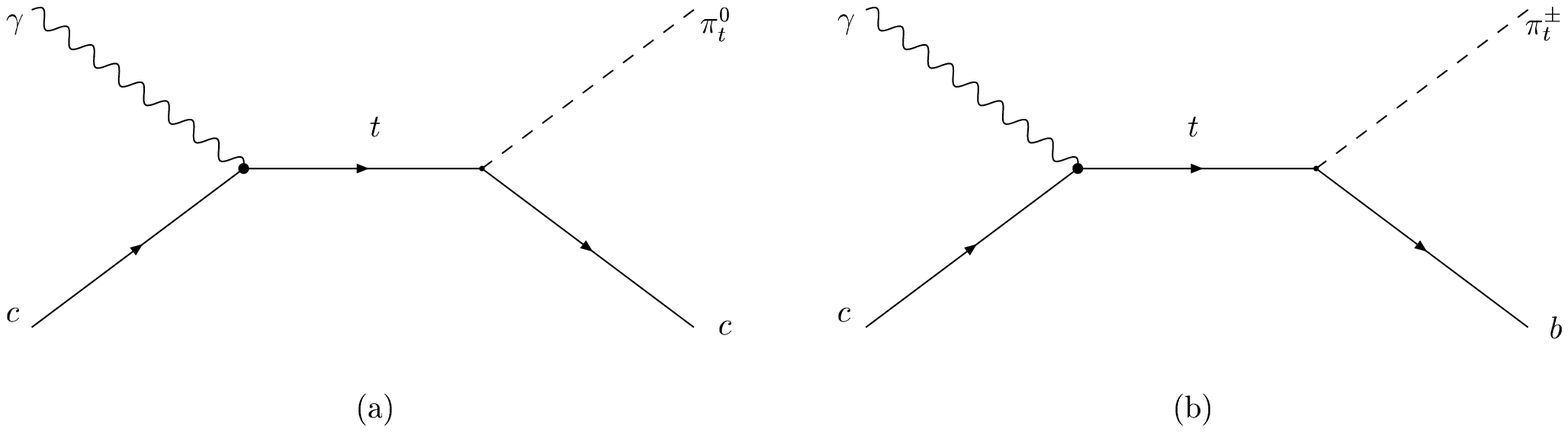,width=420pt,height=650pt} \vspace*{-10cm}
 \caption{Feynman diagrams for the top-pion production mediated by  the
 anomalous top quark couplings $tc\gamma$.}
 \label{ee}
\end{center}
\vspace*{-1.0cm}
\end{figure}
\vspace*{0cm} The production of the top-pions at the THERA
collider based $\gamma p$ collisions is mediated by the anomalous
top quark coupling $tc\gamma$ via the subprocesses $\gamma
c\rightarrow t\rightarrow \pi^{0}_{t}c$ and $\gamma c\rightarrow
t\rightarrow \pi^{\pm}_{t}b$ with the relevant Feynman diagrams
shown in Fig.1. Using the effective vertice
$\Lambda^{\mu}_{tc\gamma}$  given by Eq.(2), we can obtain the
cross section $\hat{\sigma}_{1}(\hat{s})$ and
$\hat{\sigma}_{2}(\hat{s})$ of the subprocesses $\gamma
c\rightarrow t\rightarrow \pi^{0}_{t}c$ and $\gamma c\rightarrow
t\rightarrow \pi^{\pm}_{t}b$, respectively:
\begin{eqnarray}
&&\hat{\sigma}_{1}(\hat{s})=\int_{0}^{\pi}\frac{1}{32\pi}\frac{(\hat{s}-m_{\pi_{t}}^{2})}{\hat{s}^{2}}
\overline{\sum}\mid
M_{1} \mid^{2}\sin\theta d\theta,\\
&&\hat{\sigma}_{2}(\hat{s})=\int_{0}^{\pi}\frac{1}{32\pi}\frac{(\hat{s}-m_{\pi_{t}}^{2})}{\hat{s}^{2}}
\overline{\sum}\mid
M_{2} \mid^{2}\sin\theta d\theta,
\end{eqnarray}
with
\begin{eqnarray*}
&&M_{1}=-\frac{m_{t}}{\sqrt{2}F_{t}}\frac{\sqrt{\nu^{2}_{w}-F^{2}_{t}}}{\nu_{w}}K^{tc}_{UR}K^{tt*}_{UL}
\overline{u_{c}^{\prime}}\gamma_{5}\frac{(\gamma\cdot
p_{t}+m_{t})}{\hat{s}-m_{t}^{2}+im_{t}\Gamma}\Lambda^{\mu}_{tc\gamma}u_{c}\varepsilon^{\mu},\\
&&M_{2}=i\frac{m_{t}}{F_{t}}\frac{\sqrt{\nu^{2}_{w}-F^{2}_{t}}}{\nu_{w}}K^{tt*}_{UR}K^{tc}_{DL}
\overline{u_{b}}\gamma_{5}\frac{(\gamma\cdot
p_{t}+m_{t})}{\hat{s}-m_{t}^{2}+im_{t}\Gamma}\Lambda^{\mu}_{tc\gamma}u_{c}\varepsilon^{\mu}.
\end{eqnarray*}
Where $\sqrt{\hat{s}}$ is the center-of-mass energy of the
subprocesses $\gamma c\rightarrow t\rightarrow \pi^{0}_{t}c$ and
$\gamma c\rightarrow t\rightarrow \pi^{\pm}_{t}b$ in $ep$
collisions.

The hard photon beam of the $\gamma p$ collider can be obtained
from laser backscattering at $ep$ collision in the THERA collider.
 After calculating the cross section $\hat{\sigma_{i}}(\hat{s})$ of the subprocess $\gamma
c\rightarrow t\rightarrow \pi^{0}_{t}c$ or $\gamma c\rightarrow
t\rightarrow \pi^{\pm}_{t}b$, the total production cross sections
of the neutral top-pion $\pi^{0}_{t}$ and charged top-pions
$\pi^{\pm}_{t}$ at the THERA collider can be obtained by folding
$\hat{\sigma_{i}}(\hat{s})$ with the charm quark distribution
function $f_{c/p}(x)$ in the proton $^{[19]}$ and the Compton
backscattered high-energy photon spectrum
$f_{\gamma/e}(\frac{\tau}{x})$ $^{[20]}$:
\begin{eqnarray}
\sigma(s)=\int^{0.83}_{\tau_{min}}d\tau\int^{1}_{\tau/0.83}\frac{dx}{x}f_{\gamma/e}
(\frac{\tau}{x})f_{c/p}(x)\hat{\sigma}(\hat{s}),
\end{eqnarray}
with $\hat{s}=\tau s$,
$\tau_{min}=\frac{(m_{\pi_{t}}^{2}+m_{q}^{2})^{2}}{s}$ and
\begin{eqnarray*}
&&f_{\gamma/e}(x)=\frac{1}{1.84}[1-x+\frac{1}{1-x}[1-\frac{4x}{x_{0}}(1-\frac{x}{x_{0}(1-x)})]]
\hspace{0.5cm}(x_{0}=4.83).
\end{eqnarray*}
\begin{figure}[htb]
\vspace*{-0.5cm}
\begin{center}
\epsfig{file=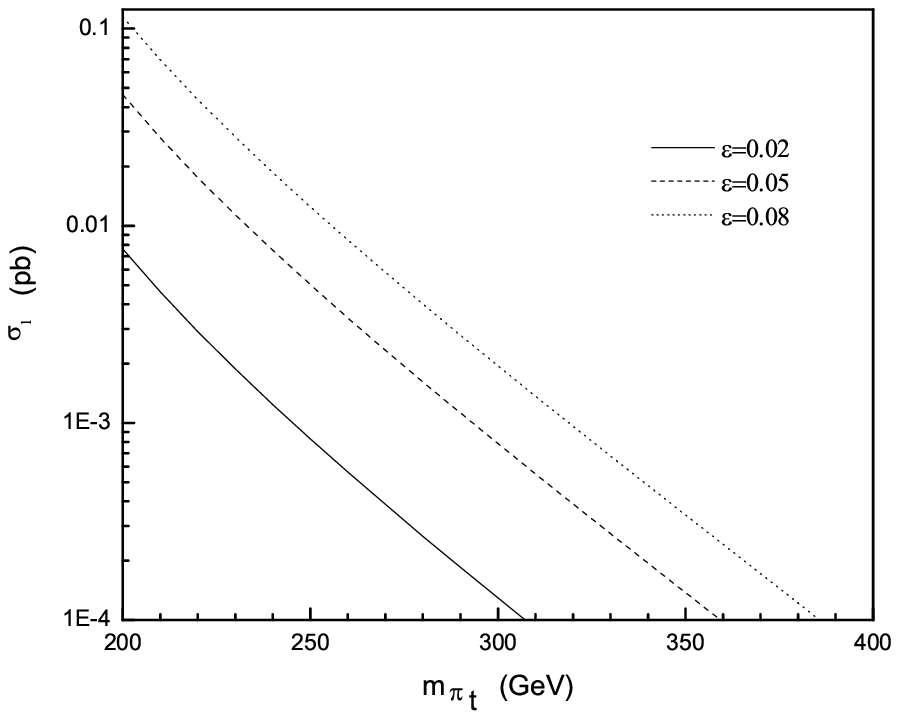,width=350pt,height=260pt} \vspace*{-1cm}
 \caption{The production cross section $\sigma_{1}$ of the process $ep\rightarrow \gamma c\rightarrow \pi_{t}^{0}c$
 as a function of $m_{\pi_{t}}$ for $\sqrt{s}=1000GeV$ and three values of the parameter $\varepsilon$.}
 \label{ee}
\end{center}
\end{figure}
\vspace*{0cm}

\begin{figure}[htb]
\vspace*{0cm}
\begin{center}
\epsfig{file=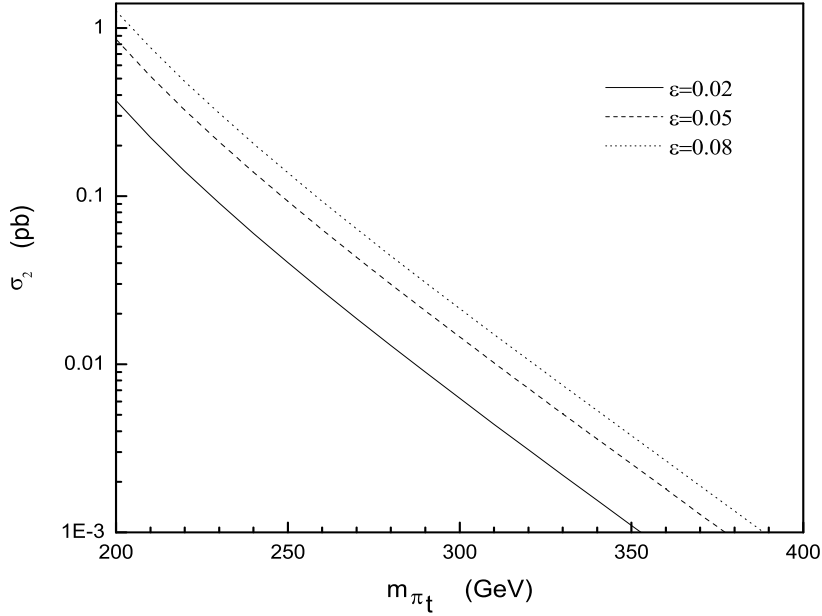,width=350pt,height=270pt}
 \vspace*{-1.cm}
 \caption{The production cross section $\sigma_{2}$ of the process $ep\rightarrow \gamma c\rightarrow \pi_{t}^{\pm}b$
 as a function of $m_{\pi_{t}}$ for $\sqrt{s}=1000GeV$ and three values of the parameter $\varepsilon$.}
 \label{ee}
\end{center}
\vspace*{-0cm}
\end{figure}

To obtain numerical results, we take the fine structure constant
$\alpha_{e}=\frac{1}{128.8}$, $m_{t}=175GeV$, $m_{c}=1.2GeV
^{[21]}$ and assume that the total decay width of the top quark is
dominated by the decay channel $t\rightarrow Wb$, which has been
taken $\Gamma(t\rightarrow Wb)=1.56GeV$. The parton distribution
function $f_{c/p}(x)$ of the charm quark runs with the energy
scale. In our calculation, we take the CTEQ5 parton distribution
function$^{[19]}$ for $f_{c/p}(x)$.

The production cross sections of the neutral top-pion
$\pi^{0}_{t}$ and the charged top-pions $\pi^{\pm}_{t}$ at the
THERA collider are plotted in Fig.2 and Fig.3, respectively, as
functions of the top-pion mass $m_{\pi_{t}}$ for
$\sqrt{s}=1000GeV$ and three values of the parameter $\epsilon$:
$\epsilon=0.02$ (solid line), $0.05 $(dash line), $0.08 $ (dotted
line). We can see that the production cross sections decrease with
$m_{\pi_{t}}$ increasing and the production cross section of
$\pi^{\pm}_{t}$ is larger than that of $\pi^{0}_{t}$ in all of the
parameter space. For $\sqrt{s}=1000GeV$, $200GeV\leq
m_{\pi_{t}}\leq 400GeV$ and $0.02\leq \epsilon \leq 0.08$, the
production cross section of the processes
$ep\rightarrow\pi^{0}_{t} c$ and $ep\rightarrow\pi^{\pm}_{t} b$
are in the ranges of $4.1\times 10^{-6}pb\sim 0.1 pb$ and $2
\times 10^{-4}pb\sim 1.2 pb$, respectively. If we assume the
yearly integrated luminosity $L=750 pb^{-1}$ for the THERA
collider based $\gamma p$ collision with $\sqrt{s}=1000GeV
^{[12]}$, then the number of the yearly production events of the
neutral top-pion $\pi_{t}^{0}$ is larger than 10 only for
$\epsilon\geq 0.08$ and $m_{\pi_{t}}\leq 220 GeV$. Thus, it is
very difficult to detect  $\pi_{t}^{0}$ via the process $e
p\rightarrow\pi^{0}_{t} c$ at the THERA based $\gamma p$
collisions. However, it is not this case for the charged top-pions
$\pi_{t}^{\pm}$. There may be several hundreds $\pi_{t}^{\pm} b$
events to be generated per year in most of the parameter space of
the TC2 models.

  It is well known that the SM is an effective theory valid only
below some high energy scale $\Lambda$, strong EWSB theories might
be needed. The strong top dynamical models, such as TC2 models,
are the modern dynamical models of EWSB. Such type of models
generally predict the existence of the top-pions. Direct
observation of these new particles via their large top Yukawa
couplings would be confirmation that the EWSB sector realized in
nature is not the SM or part of the MSSM. In this letter, we study
the production of the top-pions at the THERA  collider based
$\gamma p$ collisions in the context of the TC2 models. We find
that the top-pions can be produced via the process
$ep\rightarrow\gamma c\rightarrow \pi^{0}_{t}c$ or
$ep\rightarrow\gamma c\rightarrow \pi^{\pm}_{t}b$ mediated by the
anomalous top quark coupling $tc\gamma$, which comes from the
tree-level FC scalar couplings $\pi^{0}_{t}tc$ and
$\pi^{\pm}_{t}bc$. However, the production cross section of the
process $ep\rightarrow\gamma c\rightarrow \pi^{0}_{t}c$ is very
small. The neutral top-pion $\pi^{0}_{t}$ can not be detected via
this process at the THERA collider. For the charged top-pions
$\pi^{\pm}_{t}$, the production cross section is significantly
larger than that of $\pi^{0}_{t}$. Over a wide range of the
parameter space, there are over 100 events of $\pi^{\pm}_{t}$ to
be generated. Thus, the charged top-pions $\pi^{\pm}_{t}$ might be
detected via the process $ep\rightarrow\gamma
c\rightarrow\pi^{\pm}_{t}b$ at the THERA collider based on $\gamma
p$  collisions.

\vspace{.5cm} \noindent{\bf Acknowledgments}

  We thank Qinghong Cao for pointing out
that we should use the evolved parton distribution function of the
charm quark to calculate the production cross section. This work
was supported by the National Natural Science Foundation of China
(90203005).

\null

\end{document}